\documentclass[twocolumn,showpacs,pre,superscriptaddress]{revtex4}%
\usepackage{dcolumn}
\usepackage{amsmath}
\usepackage{graphicx}
\usepackage{amsfonts}%
\usepackage{amssymb}%

\begin{document}

\preprint{CMT-MCC1}
\title{Finite Size Scaling of Domain Chaos}
\author{M. C. Cross}
\affiliation{Condensed Matter Physics, California Institute of Technology, Pasadena CA 91125}
\author{M. Louie}
\affiliation{Applied and Computational Mathematics, California Institute of Technology, Pasadena CA 91125}
\author{D. Meiron}
\affiliation{Applied and Computational Mathematics, California Institute of Technology, Pasadena CA 91125}

\begin{abstract}
Numerical studies of the domain chaos state in a model of rotating
Rayleigh-B\'{e}nard convection suggest that finite size effects may account
for the discrepancy between experimentally measured values of the correlation
length and the predicted divergence near onset.

\end{abstract}
\date{November 23, 2000}
\maketitle

Spatiotemporal chaos is the name given to states in driven nonequilibrium
systems that are disordered in space and show chaotic time dynamics. The usual
diagnostic tools developed for chaotic dynamics in nonlinear systems with few
dynamical degrees of freedom focus on the geometrical structures in the phase
space of the dynamics. These largely lose their appeal for the very high
dimensional dynamical systems of spatiotemporal chaos. The questions of how to
characterize, understand, and predict the properties of spatiotemporally
chaotic systems remain poorly understood, despite much experimental and
theoretical attention over the past decade. Given this lack of understanding
it is important to study systems which are favorable for both experimental and
theoretical study, and in particular ones where theoretical predictions can be
made and tested experimentally. In this paper we present results that further
the comparison between theory and experiment for the state known as domain
chaos \cite{Zhong91physD} in rotating Rayleigh-B\'{e}nard convection.

Rayleigh-B\'{e}nard convection has long served as a canonical example of
pattern formation in systems far from equilibrium. A fluid confined between
two horizontal plates and driven thermally by maintaining the bottom plate at
a higher temperature than the top plate undergoes an instability to a state in
which there is fluid motion driven by the buoyancy forces induced by the
thermal expansion. Far away from lateral boundaries the structure of the fluid
motion locally forms the familiar convection rolls with a diameter close to
the depth of the cell. This spontaneous formation of spatial structure in a
uniformly driven system is known as pattern formation.

The instability to the roll state occurs at a particular value the Rayleigh
number (a dimensionless measure of the temperature difference across the
fluid) $R=R_{c}$. For values of $R$ slightly above $R_{c}$ the pattern is
usually found to be a time independent state. However if the convection
apparatus is now rotated about a vertical axis, Coriolis effects perturb the
fluid velocity, and above a critical rotation rate $\Omega_{c}$ an ideal
pattern of straight convection rolls is predicted to become unstable via the
``Kuppers-Lorz'' instability \cite{Kuppers69}. The nature of the instability
is that the rolls become unstable towards the growth of a second set of rolls
with their axis rotated by an angle $\theta_{KL}$ from the axis of the
original set of rolls. The value $\theta_{KL}$ depends on fluid properties,
but for typical fluids is close to $60%
{{}^\circ}%
$ at the onset of the instability. Once the second set of rolls grows to
saturation replacing the first set, they in turn will become unstable towards
rolls rotated through a further $\theta_{KL}$, etc. It is predicted that there
will be no time independent saturated state even arbitrarily close to onset.
Experimentally the state of domain chaos is found, consisting of domains of
differently oriented rolls with a persistent dynamics of domains expanding at
the expense of one (or more)\ of the neighboring domains through the motion of
the domain walls between them.

The domain chaos state is of particular interest since it survives down to the
onset of the convective state, where theoretical treatment based on an
expansion in the weak nonlinearity should be possible. Indeed, based on the
approximation that the switching angle is exactly $60%
{{}^\circ}%
$, Cross and Tu \cite{Tu92} developed a simple model of the domain chaos
state, extending earlier work of Busse and Heikes \cite{Busse80a}. The model
is for the coupled dynamics of domains of rolls at only three orientations,
characterized by three amplitudes $A_{i}(x,y,t)$, $i=1,3$ giving the strengths
of the three roll components at each point in space and time. The coupled
amplitude equations they used (after appropriate rescaling of space and time
coordinates to eliminate unimportant constants) take the form
\begin{subequations}
\label{Eq_Amplitude}%
\begin{align}
\partial_{t}A_{1}  &  =\varepsilon A_{1}+\partial_{x_{1}}^{2}A_{1}-(A_{1}%
^{2}+g_{+}A_{2}^{2}+g_{-}A_{3}^{2})A_{1},\\
\partial_{t}A_{2}  &  =\varepsilon A_{2}+\partial_{x_{2}}^{2}A_{2}-(A_{2}%
^{2}+g_{+}A_{3}^{2}+g_{-}A_{1}^{2})A_{2},\\
\partial_{t}A_{3}  &  =\varepsilon A_{3}+\partial_{x_{3}}^{2}A_{3}-(A_{3}%
^{2}+g_{+}A_{1}^{2}+g_{-}A_{2}^{2})A_{3}.
\end{align}
Here the variable $x_{i}$ is the spatial coordinate in the direction
perpendicular to the $i$th set of rolls; the derivatives transverse to these
directions are of higher order. Although in general complex amplitudes should
be used, Cross and Tu made the simplification of assuming the $A_{i}$ to be
real. This corresponds to neglecting variations of the wave numbers of the
rolls. The parameter $\varepsilon$ measures the distance from onset
($\varepsilon=(R-R_{c})/R_{c}$), and is the small parameter of the expansion.
The constants $g_{+}$ and $g_{-}$ determine the interaction between one set of
rolls and the set rotated through $+60%
{{}^\circ}%
$ and $-60%
{{}^\circ}%
$ respectively. In the absence of rotation, clockwise and anticlockwise
rotations are equivalent so that $g_{+}=g_{-}$ and in this limit it is easily
shown that Eqs.(\ref{Eq_Amplitude}) are relaxational or ``potential''
\cite{Cross93} so that persistent dynamics is impossible. The rotation breaks
this ``chiral symmetry'', so that $g_{+}\neq g_{-}$, and the equations are
then no longer relaxational. Cross and Tu found numerically, for sufficiently
different $g_{+},g_{-}$, a state of persistently dynamic domains.

The demonstration of the \emph{existence} of the chaotic domain state within
Eq.~(\ref{Eq_Amplitude}) gives immediate \emph{quantitative} predictions for
the scaling of the size and lifetime of the domains with $\varepsilon$.
Defining scaled variables $X_{i}=\varepsilon^{1/2}x_{i}$, $T_{i}=\varepsilon
t$, and $\bar{A}_{i}=\varepsilon^{-1/2}A_{i}$, yields equations with no
appearance of the parameter $\varepsilon$. In these scaled variables the
domain size and lifetime are therefore $\varepsilon$ independent, leading to
the prediction of a domain size or correlation length $\xi$ of the state
scaling as $\varepsilon^{-1/2}$ and a domain lifetime or correlation time
$\tau$ of the dynamics scaling as $\varepsilon^{-1}$ in the physical
variables. This remains one of the few quantitative predictions for properties
of spatiotemporal chaos in an experimental dissipative system far from
equilibrium that has actually been tested experimentally. However recent
experiments \cite{Hu95prlrotation} did not find the predicted result. It is
this important discrepancy that we address in the present paper.

The clear disagreement between the experiment and the theory is that in the
experiment $\xi^{-2}$ and the domain switching rate $\omega_{a}=$ $\tau^{-1}$
appear to remain finite as $\varepsilon$ approaches zero, instead of going to
zero linearly in $\varepsilon$. If a power law dependence on $\varepsilon$ is
forced on the data, powers much smaller than $1$ result. Based on numerical
simulations of equations showing domain chaos we suggest that this discrepancy
between experiment and theory might be due to the (necessarily) finite size of
the experimental system.

Our conclusion is based on the following. We find results for the correlation
length in numerical simulations that have many of the same qualitative
features as in the experiment. In particular the measured correlation length
$\xi_{M}$ (using an algorithm defined below that is the same as the one used
in the experimental work) appears to remain finite approaching the threshold.
The flexibility of the numerical approach allows us to investigate this
behavior as a function of the aspect ratio $\Gamma$. We find that the data for
different $\Gamma$ and different $\varepsilon$ collapse onto a single form
suggested by finite size scaling
\end{subequations}
\begin{equation}
\xi_{M}=\xi f(\xi/\Gamma) \label{Eq_Scaling}%
\end{equation}
with $\xi\propto\varepsilon^{-1/2}$ the ``ideal'' correlation length following
the theoretical prediction. Our numerical data is consistent with
$f(x)\rightarrow const$ for small $x$ so that in large enough systems (or for
large enough $\varepsilon$) the predicted dependence proportional to $\xi
\sim\varepsilon^{-1/2}$ would be found, whereas $f(x)\propto x^{-1}$ for large
$x$, so that in small systems (or for small $\varepsilon$) the measured
correlation length is proportional to the system size.

The equation we simulate is the partial differential equation for a real field
$\psi(x,y,t)$ in a two dimensional domain%
\begin{multline}
\partial_{t}\psi=\varepsilon\psi+(\nabla^{2}+1)^{2}\psi-g_{1}\psi
^{3}\label{Eq_SH}\\
+g_{2}\mathbf{\hat{z}\cdot}\nabla\times\lbrack(\nabla\psi)^{2}\nabla
\psi]+g_{3}\nabla\mathbf{\cdot}[(\nabla\psi)^{2}\nabla\psi].
\end{multline}
with boundary conditions%
\begin{equation}
\psi=\mathbf{\hat{n}\cdot\nabla\psi=0} \label{Eq_BC}%
\end{equation}
on the boundary with normal $\mathbf{\hat{n}}$. This equation has been
introduced previously to model domain chaos \cite{CrossMeironTu94} with
periodic boundary conditions. With $g_{2}=g_{3}=0$ it is the well known
Swift-Hohenberg equation that has been much studied in the context of pattern
formation \cite{Cross93}. The spatially uniform state $\psi=0$ becomes
unstable to a stripe state with wave number $q_{c}=1$ at $\varepsilon=0$, and
for $\varepsilon>0$ steady, stable, nonlinearly saturated stripe states may be
found. In modelling a convection system we can imagine $\psi(x,y)$ as
representing the temperature field across a midplane of the system, and the
stripes are a section of the convection rolls. The second nonlinear term in
Eq.~(\ref{Eq_SH}), with coefficient $g_{2}$ is the all important term yielding
the breaking of the chiral symmetry induced by rotation in the physical
system---and so increasing $g_{2}$ corresponds to increasing rotation rate in
the convection system. For sufficiently large $g_{2}$ the stripe state becomes
unstable to a rotated set of stripes as in the Kuppers-Lorz instability. The
angle at which the new set of stripes occurs can be tuned with the parameter
$g_{3}$ \cite{CrossMeironTu94}.

Although Eq.~(\ref{Eq_SH}) is easier to evolve numerically than the coupled
equations for fluid velocity and temperature fields in a three dimensional
domain that give a complete description of the convection system, the task
remains challenging. This is because we need to integrate the equation in a
large domain and over long times. In fact, since we are interested in
approaching the limit $\varepsilon\rightarrow0$ to uncover the scaling
behavior, and, in this limit, the dynamics becomes very slow, exceedingly long
integration times are necessary. In addition, to model the experiment the
domain must be circular, and to eliminate any bias towards a particular stripe
orientation the use of circular polar coordinates to describe the geometry
seems preferable.

To meet these numerical challenges we have developed a \emph{fully implicit}
method using a finite difference representation of the differential operators
on a polar coordinate mesh. At each time step the nonlinear Crank-Nicolson
equations for the new value of the field are solved by Newton's method. The
GMRES (generalized minimal residual) iterative method \cite{Edwards94},
preconditioned with the Bjorstad \cite{Bjorstad83} fast direct biharmonic
solver, is used to solve the linear problem within each Newton iteration. This
method allowed large time steps $\Delta t$, for example up to $100$ at
$\varepsilon=0.01$. A variable time stepping algorithm was implemented to
exploit this opportunity, based on a comparison of results with time steps
$\Delta t$ and $\Delta t/2$. This time stepping should be compared with what
might be obtained with a conventional semi-implict method where the gradient
terms in the nonlinear terms impose a stability limit on the possible time
step determined by the spatial mesh resolution. This limitation is
particularly severe using polar coordinates since the mesh spacing in the
azimuthal direction becomes very small near the polar origin. In practice we
found a time step limited to $\Delta t<0.1$ at $\varepsilon=0.01$ using a
conventional semi-implicit code, a factor of $1000$ smaller than in the fully
implicit code.

The polar mesh induces an artificial singularity in the description at the
origin of the circle where the physical behavior will be smooth. This
difficulty has plagued the development of many codes based on polar
coordinates. To test the accuracy of our code in the vicinity of the origin we
simulated Eq.~(\ref{Eq_SH}) with $g_{1}=g_{2}=0$ starting with an initial
condition of straight stripes. With these parameters the stripe state is
unstable to a square pattern. For an initial condition that is exactly a
stripe state, the development of squares can physically only be initiated at
the boundary, and the square pattern should then be observed to propagate in
from the boundaries. With inadequate spatial resolution we found instead that
squares also began to grow via nucleation around the origin, presumably an
artifact of the reduced accuracy of the numerical integration here. This
unphysical result could be effectively eliminated by increasing the number of
radial mesh points (as mentioned above, the azimuthal resolution for a fixed
number of azimuthal points becomes very fine near the origin). To take the
best advantage of the computer resources we used a \emph{variable} mesh in the
radial direction, with the resolution increasing smoothly to a factor of two
or three improvement over the inner third of the radial direction. It should
be noted that these test runs use the exponential amplification due to a
physical instability (stripes to squares) acting over a long time (e.g. a time
of 100) to enhance the effects of numerical inaccuracies to visible
amplitudes. Thus the elimination of visible spurious effects near the origin
in these test runs gives us confidence in the accuracy of the numerical procedure.%

\begin{figure}
[tbh]
\begin{center}
\includegraphics[
height=4.0in
]%
{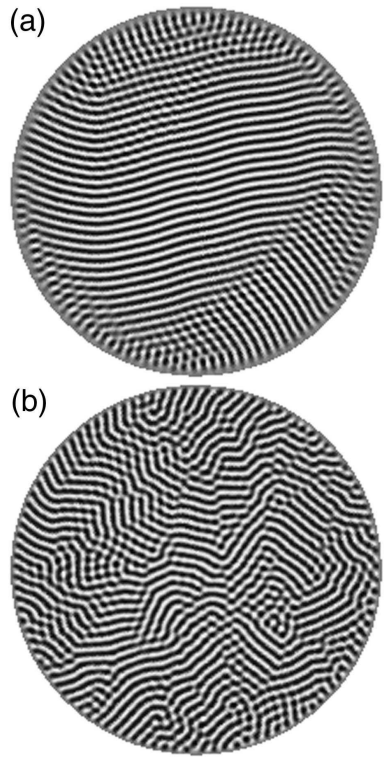}%
\caption{Snapshots of the dynamic domain configuration at two values of the
control parameter (a) $\varepsilon=0.01$ and (b) $\varepsilon=0.3$. The aspect
ratio is $40\pi$.}%
\label{Domains}%
\end{center}
\end{figure}
To study domain chaos we simulated Eq.~(\ref{Eq_SH}) with fixed values of the
nonlinear coefficients $g_{1}=1$, $g_{2}=-2.60$, and $g_{3}=1.5$ corresponding
to a Kuppers-Lorz instability at an angle of $60%
{{}^\circ}%
$ \cite{CrossMeironTu94}. We studied the behavior for values of $\varepsilon$
between $0.01$ and $0.3$ and in circular geometries with radii $30\pi$,
$40\pi$, $50\pi$, and $80\pi$, corresponding to aspect ratios of $30-80$ in
the experimental system. Snapshots of the field $\psi(x,y)$ from two runs are
shown in Fig.~(\ref{Domains}). Notice the dependence of the domain size on the
control parameter $\varepsilon$, and the comparable values of domain and
system sizes at the smaller value of $\varepsilon$.%

\begin{figure}
[tbh]
\begin{center}
\includegraphics[
height=2.5304in,
width=3.4126in
]%
{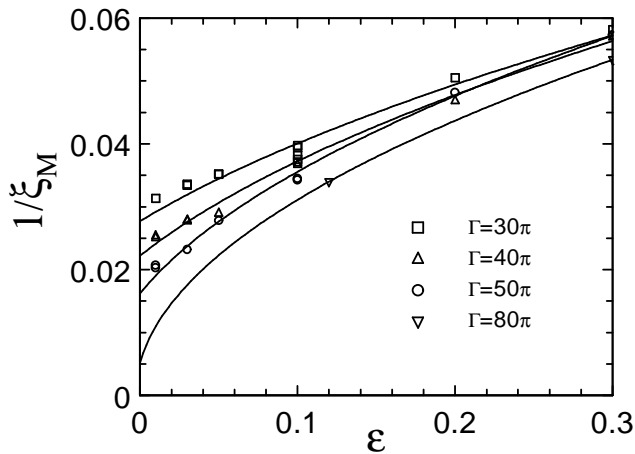}%
\caption{Inverse measured correlation length $\xi_{M}^{-1}$ as a function of
control parameter $\varepsilon$ for various aspect ratios $\Gamma$. The lines
are fits of the form $\xi_{M}^{-1}=a(\varepsilon-\varepsilon_{0})^{1/2}$.}%
\label{Fig_Lengths}%
\end{center}
\end{figure}
The quantitative diagnostics of the state are based on the Fourier transform
$\tilde{\psi}(\mathbf{k)}$ using a Hanning window over an inscribed square,
following the same approach as in the experiments \cite{Ning93pre2}. The
intensity $S(\mathbf{k})=|\tilde{\psi}(\mathbf{k})|^{2}$ is found to be
concentrated on a ring in Fourier space near $k=1$ corresponding to the stripe
periodicity of $2\pi$. The intensity varies with the angle around this ring,
corresponding to the varying representation of domains of different stripe
orientations in the pattern. This angular distribution varies with time as the
domains evolve, and can be used to define the switching rate $\omega_{a}$. The
correlation length of the pattern is defined as the average width of the ring
in Fourier space $\xi_{M}=[\left\langle k^{2}\right\rangle -\left\langle
k\right\rangle ^{2}]^{-1/2}$ with
\begin{equation}
\left\langle k^{n}\right\rangle =\left\langle \frac{\int k^{n}S(\mathbf{k}%
)d^{2}k}{\int S(\mathbf{k})d^{2}k}\right\rangle _{t}%
\end{equation}
where $\left\langle \,\right\rangle _{t}$ denotes a time average.%

\begin{figure}
[tbh]
\begin{center}
\includegraphics[
width=3.35in
]%
{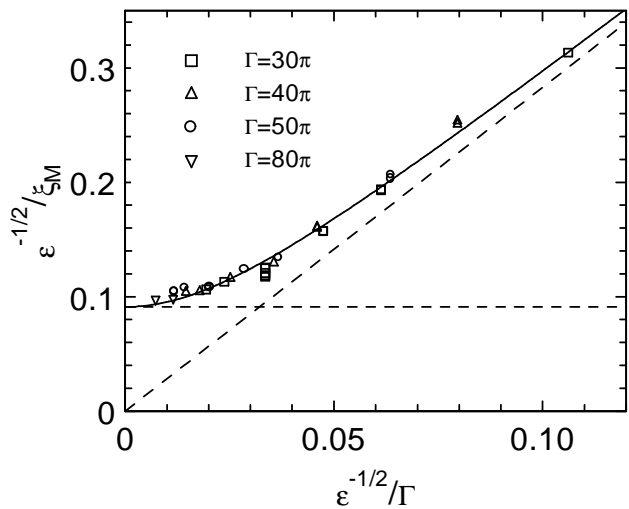}%
\caption{Plot of the inverse correlation length $\xi_{M}^{-1}$ scaled with
$\varepsilon^{1/2}$ against the inverse aspect ratio $\Gamma^{-1}$ scaled by
$\varepsilon^{-1/2}$ showing the collapse of data for different $\varepsilon$
and $\Gamma$ onto a single curve. The solid line is a fit to the empirical
form $y=\sqrt{a^{2}+b^{2}x^{2}}$ yielding $a=0.091$ and $b=2.8$. The dashed
lines show the asymptotic limits of this fit.}%
\label{Fig_Scaling}%
\end{center}
\end{figure}
Results for the correlation length are shown in Fig.~(\ref{Fig_Lengths}). As in
the experiments the inverse of the measured correlation length $\xi_{M}^{-1}$
appears to remain nonzero at onset, and the data are reasonably well fit by a
form $\xi_{M}^{-1}\sim\sqrt{\varepsilon-\varepsilon_{0}}$ rather than the
expected $\sqrt{\varepsilon}$ dependence. The dependence of $\varepsilon_{0}$
we find on the aspect ratio suggests that this might be understood as a finite
size effect. This is confirmed by the scaling plot motivated by
Eq.~(\ref{Eq_Scaling}) which suggests that a plot of $\varepsilon^{-1/2}%
/\xi_{M}$ against $\varepsilon^{-1/2}/\Gamma$ should collapse the data. The
successful collapse is shown in Fig.~(\ref{Fig_Scaling}). Note that for small
values of $\varepsilon^{-1/2}/\Gamma$ (i.e. large $\Gamma$ or $\varepsilon$)
the value of $\xi_{M}\varepsilon^{-1/2}$ approaches a constant corresponding
to the theoretical expectation for large enough systems. On the other hand for
large $\varepsilon^{-1/2}/\Gamma$ the curve approached a linear behavior
consistent with the correlation length scaling simply with the aspect ratio,
$\xi_{M}\simeq\Gamma/2.8$. Note that the finite size corrections become
important (e.g. identified as the intersection point of the straight lines in
Fig.~(\ref{Fig_Scaling}) for $\Gamma\leq3\xi$, with an additional factor of $3$
over the naive expectation.

In the experimental work there has been no systematic attempt to study the
dependence on aspect ratio, which is much harder to do experimentally than
numerically. The experiments did investigate the dependence of the correlation
length on the rotation rate, and found that the measured correlation lengths
at different rotation rates could be related by an $\varepsilon$-independent
scale factor: $\xi(\varepsilon,\Omega)\simeq\xi_{0}(\Omega)f(\varepsilon)$.
Since the deviation from the expected $\varepsilon^{-1/2}$ behavior found in
$f(\varepsilon)$ is common to all the runs at different $\Omega$, this scaling
does not shed light on the basic discrepancy with the predicted $\varepsilon
^{-1/2}$ scaling however.

We have also studied the behavior of the switching frequency $\omega_{a}$ on
$\varepsilon$. Here, unlike the experiment, we find good agreement with the
prediction $\omega_{a}\propto\varepsilon$ for small $\varepsilon$ with no
dependence on the aspect ratio. The trend $\omega_{a}\rightarrow0$ for
$\varepsilon\rightarrow0$ even in finite size systems is not surprising for
Eqs.(\ref{Eq_SH},\ref{Eq_BC}) since the effects of the rotation that lead to
the persistent dynamics \emph{only appear in the nonlinear terms} which become
small for $\varepsilon\rightarrow0$. A careful analysis of the fluid equations
shows that there are also \emph{linear} terms depending on the rotation that
are important near the boundaries. This means that in a finite system the
onset of convection rolls in a rotating system is actually a Hopf bifurcation
with the onset frequency going to zero for large aspect ratios, and so it
might not be surprising that $\omega_{a}$ remains nonzero at onset. These
effects can in fact be captured using modified boundary conditions for the
Swift-Hohenberg equations \cite{Kuo94}. However these complicated boundary
conditions make the numerical scheme considerably harder, and we have not
modified our code to investigate them.

In conclusion our numerical simulations of model equations for rotating
Rayleigh-B\'{e}nard convection show results for the correlation length near
threshold that show qualitative similarities to the experimental results. A
finite size scaling ansatz shows that our measured lengths are in fact
consistent with the expected $\varepsilon^{-1/2}$ divergence near threshold,
but that the finite system size obscures this dependence. It will be
interesting to see if this result can be confirmed experimentally, or in
simulations of the full fluid and heat equations for convection that we are
currently pursuing. It is noteworthy that similar predictions for a diverging
correlation time near threshold in electroconvection spatiotemporal chaos have
recently been verified experimentally \cite{Toth98}. In this system larger
aspect ratios are accessible so that finite size effects should not be important.

\begin{acknowledgments}
This work was partially supported by the Division of Materials Science and
Engineering of Basic Energy Sciences at the Department of Energy, Grant DE-FG03-98ER14891
\end{acknowledgments}


\end{document}